# Strong g-Factor Anisotropy in Hole Quantum Dots Defined in Ge/Si Nanowires

*Stefano Roddaro, Andreas Fuhrer\*, Carina Fasth, Lars Samuelson*

Solid State Physics/the Nanometer Structure Consortium, Lund University, Box 118, S-221 00, Lund, Sweden

*Jie Xiang and Charles M. Lieber*

Department of Chemistry and Chemical Biology, Harvard University, Cambridge MA 02138, US

\*E-mail: fuhrer@nigra.ch

**Abstract** We demonstrate fully tunable single and double quantum dots in a one-dimensional hole system based on undoped Ge/Si core-shell nanowire heterostructures. The local hole density along the nanowire is controlled by applying voltages to five top gate electrodes with a periodicity of 80 nm, insulated from the wire by a 20 nm-thick $HfO_2$ dielectric film. Low-temperature transport measurements were used to investigate the magnetic field dependence of Coulomb blockade peaks in a single quantum dot and indicate a strongly anisotropic g-factor with $|g_\parallel| = 0.60 \pm 0.03$ and $|g_\perp| < 0.12$.

Nanoparticle-mediated growth of nanowires (NWs) has been at the center of a significant experimental effort in the past years[1,2] and is leading to the development of new classes of devices[3–12] that are receiving growing attention both as research tools and for their potential applications. Ge/Si coaxial NWs constitute a special and so far unique case, as they support a high-mobility one-dimensional hole system with a mean free path exceeding 170 nm even at room temperature, as suggested by recent experimental results[4–6]. The adoption of hole systems is also expected to have a strong impact on the transport characteristics in regard of the spin degrees of freedom: the intrinsic strong correlation between spin and orbital motion for holes in the valence bands typically leads to a large anisotropy of the g-factor depending on the real-space geometry and on quantum confinement[13–17]. In the specific case of a cylindrical wire it has been predicted that the axial g-factor $g_\parallel$ should depend on the axial angular momentum state while $g_\perp$ should be strongly suppressed[18]. The study of these spin-orbit related properties of one-dimensional holes will be crucial for assessing the viability of quantum electronic devices based on Ge/Si NWs. In this Letter we report a marked anisotropy in the g-factor for holes and we discuss this result in light of available theories[14,18]. Our experiments also demonstrate the viability of tunable multi-dot systems in these Ge/Si NWs.

Our Ge/Si NWs were grown in a two-step chemical vapor deposition process: in the first step Ge NWs with a diameter of 15 ± 4 nm were fabricated by gold nano cluster assisted catalytic growth; in the second step the NW surface was passivated by overgrowing a Si shell with a thickness of 1.5 to 2 nm. The valence band offset between Ge and Si ($\Delta E_V \approx 500$ mV) leads to a natural population of holes in the Ge core so that intentional doping can be avoided as a result of which transport properties are greatly improved. Further details on the growth and NW structure have been reported previously[5]. NWs were dispersed in ethanol from the growth substrates by sonication and deposited on a degenerately doped Si substrate with a 600 nm-thick thermal oxide layer. They were then located in relation to predefined markers using an scanning electron microscope and contacted by electron beam lithography, using annealed Ni ohmic contacts[5] separated by an 800 nm gap. A $HfO_2$ dielectric with a thickness of 25 ± 5 nm was deposited over the NWs and contacts, and then seven 50 nm thick Ti/Au gate fingers were placed on top of the encapsulated structure as shown in Fig. 1(a). Five of these gate electrodes $g_1 - g_5$ [blue in Fig. 1(a)] were used to locally tune the band edge positions to form tunable tunnel barriers and adjust the hole density within the wire.

These gates were 50 nm wide with a periodicity of 80 nm and were aligned to lie within the gap between the ohmic contacts (yellow). Two larger gates $g_S$ and $g_D$ (red), partially overlapping the ohmic contacts, were kept at a large negative voltage down to −5.0 V in order to maintain a high hole density and thus ensure good carrier injection into the NW even for large positive voltages on the inner gate electrodes. Local gates $g_1 - g_5$ were either biased to positive voltages above 0.6 V to define barriers along the wire or to bias values close to 0 V when used as plunger gate electrodes to tune the number of charge carriers in the quantum dots. Transport measurements were performed at 4.2 K by dipping the sample into a liquid helium dewar and in a $^3$He cryostat with a base temperature of 250 mK.

By tuning the five gate voltages $V_{g1}$ -$V_{g5}$ a variety of dot configurations along the NW can be obtained. We start our discussion with the simplest situation of a single quantum dot induced by biasing gates $g_1$ and $g_3$ close to pinch-off (0.63 V and 0.90 V respectively) and using gate $g_2$ as a plunger gate in the range 0 – 70 mV. Figure 1(b) shows a measurement of the current through the NW as a function of plunger gate voltage $V_{g2}$ at a temperature T = 250 mK, exhibiting pronounced Coulomb blockade oscillations with an average separation of $\Delta V_{g2} \approx 10$ mV. Finite bias measurements were performed using symmetric biasing conditions with $V_S = +V_b/2$ and $V_D = -V_b/2$. Figure 1(c) shows the corresponding Coulomb diamonds in the differential conductance through the NW. From these measurements we estimate a charging energy $E_C = e^2/C_\Sigma = 3$ meV and obtain the plunger gate lever-arm $\alpha_{g2}$ by taking the difference between the two borderline slopes crossing where the diamonds touch at $V_b = 0$ [see white lines in Fig. 1(c)]. This yields $\Delta V_{g2}/\Delta V_b \approx 1.68$ (dashed lines) and $\Delta V_{g2}/\Delta V_b \approx -1.44$ (solid lines) and thus results in $\alpha_{g2} \approx 0.32$. Within a simple capacitance model [see Fig. 1(d)] the gate capacitance can be evaluated as $C_{g2} = e/\Delta V_{g2} \approx 16$ aF while the corresponding value for the total capacitance is $C_\Sigma = C_{g2}/\alpha_{g2} \approx 50$ aF. From the periodicity of the Coulomb oscillations in $V_{g2}$ versus that in $V_{g1}$ and $V_{g3}$ (data not shown) we can deduce in a similar way that $C_{g1} \approx C_{g3} \approx 8$ aF. This allows us to verify that the dot is located in the NW close to central gate $g_2$ and at a similar distance from g1 and g3. The remaining capacitance ($C_\Sigma - C_{g1} - C_{g2} - C_{g3} \approx 18$ aF) is attributed to the coupling of the dot to the S and D sections of the device as well as to other remote metal electrodes and the Si substrate.

In order to determine the hole g-factor in these one-dimensional quantum dots, we study the parametric evolution of the Coulomb blockade peaks as a function of an applied magnetic field *B*. While we would have preferred to measure $g_\parallel$ and $g_\perp$ separately by aligning the field to the NW axis and perpendicular to the NW, the random distribution of the NWs yielded devices where the field was tilted by an arbitrary angle $\Theta$ with respect to the NW axis. Here we report data for a device with $\Theta = 45°$ (*case i*), and in a second cool-down data from the same device with the field perpendicular to the substrate, $\Theta = 90°$ (*case ii*). Using these two configurations we determine both $g_\parallel$ and $g_\perp$.

From the peak spacing and the pinch-off value of gate $g_2$ $V_{g2} \approx 2.0$ V we estimate the number of holes on the dot to be N ≈ 200. Despite this rather large number of charge carriers we find a significant fluctuation of $\Delta\mu$ at zero magnetic field with a standard deviation σ = 0.2 meV, indicating that quantum confinement and/or interaction effects beyond the constant interaction model are important[19]. While the overall evolution of the $\Delta\mu$ curves is complex, as expected for a many-hole quantum dot with a small average level spacing ΔE, we can identify a limited number of recurrent slopes ≈ ±25 μeV/T as well as ≈ 0 μeV/T in Fig. 2(b). This is particularly evident looking at curves *II-III-IV*, where two clearly complementary slopes are visible. If we restrict ourselves to these two slopes plus a constant line we can fit all of the inter-peak distances except for the section in curve *I* between 7 T and 10 T where a charge rearrangement influences the evolution of the peak at $V_{g2}$ ≈ 10 mV [See also Fig. 2(a)]. This is evidenced also by the larger deviation between the positive and negative magnetic field data for this peak in Fig. 2 (b).

Before proceeding with our analysis it is important to consider the physical origin of the relative peak shifts we report in relation to possible orbital effects. We argue here that, given the small core (15 nm) of our Ge/Si wires, diamagnetic shifts of the orbital states do not play an important role for the inter-peak distance shown in Fig. 2 (b). For a magnetic field perpendicular to the nanowire we expect orbital energies only to be substantially modified when the magnetic length $l_m = h/eB$ becomes smaller than the expectation value for the width of the wavefunction in the strong confinement direction[20,21]. We estimate that for the investigated magnetic field range this effect is small on the scale of the experimentally observed peak shifts. Even for higher magnetic fields, diamagnetic effects will not influence the inter-peak distance $\Delta\mu$ as long as consecutive holes are filled into the same 1D subband of the wire. The corresponding consecutive Coulomb blockade peaks would all experience the same diamagnetic shifts and the peak separation would thus not be affected. *Case ii* could in principle be more delicate as the field might couple directly to states with a well-defined axial angular momentum. However, in a real device, rotational invariance of the confinement will be strongly perturbed by the one sided gating of the structure and we expect orbital levels to be rather stiff against the external field even in this case. Building on this argument, we discuss the data of Fig. 2(a) and (b) in terms of an effective spin splitting.

For magnetic fields |B| < 8 T the slopes in the yellow band in Fig. 2(b) then have a simple interpretation in terms of a sequential addition of holes with spin ↑↑↓↑↑↓ as indicated by the red arrows. In the absence of level crossings and assuming that the magnetic field dependence of the orbital levels varies slowly with hole number the level separation $\Delta\mu(B)$ is given by $\Delta\mu(B = 0) + g^*\mu_B B\Delta S$ where $\mu_B$ is the Bohr magneton and $\Delta S = 0, \pm 1$. The slopes in the low-field region of Fig. 2(b) thus determine the sequence of holes with S = 1/2 successively filling dot levels A through F. We note that the spin filing sequence is not simply alternating between ↑ and ↓ which indicates that spin ground states of the dot with S > 1/2 exist. Given the degeneracies in the valence band and the high hole number this is not unexpected. Fig. 2(c) schematically depicts the expected peak -evolution if orbital effects are neglected. For B > 7.5 T the slopes in Fig. 2(b) begin to change significantly and the different $\Delta\mu$'s undergo correlated kinks. We attribute this to two spin induced level crossings between levels B and C and between E and F. Given the slopes we determined in Fig. 2(b), the two level crossings at B = 7.6 T and B = 10 T would require hole states with a level spacing ΔE of ≈ 190 and ≈ 250 μeV respectively which is within the limits given by the magnitude of the fluctuations σ at B = 0 from our previous estimate.

In order to compare the data in Fig. 2(b) with *case ii* where the field was perpendicular to the NW, we shift the level separations $\Delta\mu$ to a common origin at B = 0 and plot the average of the two values measured at positive and negative fields in Fig. 2(d) for |B| < 8 T. From a linear fit of curves II, III and V we find a slope 24.7 ± 0.9 μeV/T corresponding to an effective g-factor

$$|g^*| = \sqrt{g_\parallel^2 \cos^2\Theta + g_\perp^2 \sin^2\Theta} = 0.43 \pm 0.02. \quad (1)$$

The same slopes occur in other similar single-dot configurations at low magnetic fields, even if noise and drifts in the data do not always allow as clear an estimate as in Fig. 2(d). Two additional curves from another dataset are plotted in Fig. 2(d) for comparison (hollow triangles).

We now move to *case ii* where the magnetic field was perpendicular to the wire. Theory for confined one-dimensional hole systems predicts a strong anisotropy of the g-factor and low values for $g_\perp$[14,18]. Our measurements are consistent with this expectation. This is particularly visible in Fig. 2(e) where we plot $\Delta\mu(B)$ for a dataset containing five consecutive Coulomb blockade peaks. Data points for positive and negative fields were consistent and their average values are plotted in the figure. The curves do not show any clear trends and linear fits performed over the single curves in the vicinity of $B = 0$ yield a maximum slope of $\approx 7$ $\mu$eV/T. This gives an estimate $|g_\perp| < 0.12$. Due to the rather unstable behavior of the device during the cool-down for $\Theta = 90°$, statistics were insufficient to warrant extracting a more precise value for $|g_\perp|$. On the other hand, there is no indication of higher g-factor values for this field orientation and using $|g_\perp| < 0.12$ in Equation (1) we estimate $|g_\parallel| = 0.60 \pm 0.03$. A more refined investigation of spin in these hole quantum dots would be possible by tuning the dots to smaller hole filling, ideally down to the last holes. The investigated devices did not allow doing this. We found that it was not possible to deplete the dot with $V_{g2}$ while staying in the single-dot regime.

It was possible, however, in the present devices to continuously tune the NW between a single dot and a configuration where two separate dots are formed in the gaps between gates $g_1/g_2$ and $g_2/g_3$. Figure 3 shows measurements of stability diagrams of the current through the wire versus both barrier gate voltages $V_{g1}$ and $V_{g3}$ in a second device with the same gate geometry to the one discussed above. Tuning $V_{g2}$ more and more negative leads to a transition from single combined dot behavior as shown in Fig. 3(a) to the typical hexagon stability diagrams expected for double quantum dots e.g. in Fig. 3(c). In Fig. 3(a) for $V_{g2} = 1.45$ V Coulomb blockade oscillations can be tuned with either of the two barrier voltages $V_{g1}$ and $V_{g3}$ giving roughly equal lever-arms for each of the two gates (see the 45° slope for the Coulomb peaks shown in black). In Figs. 3(b),(c) and d the central gate is tuned more positive ($V_{g2} = 1.55$ V, 1.85 V and 2.20 V respectively) splitting the initial dot into two separate dots with less and less coupling between the two. Looking at the slope of the white lines in the plots[22], it is possible to determine $C_{g1} \approx 8.5$ aF $C_{g3} \approx 7.5$ aF, while the inter-dot coupling can be quantified by $C_m \approx 4.0$ aF and 2.5 aF for the two intermediate cases of Fig. 3(b) and 3(c). Exploiting all available gate-electrodes in our devices, other double dot configurations, can be realized: Fig. 3(e) shows the stability diagram at $T = 250$ mK corresponding to a double dot configuration where two plunger gates control the filling in the two dots while the three barriers defining the double dot system are adjusted using other independent gate electrodes.

In conclusion we have demonstrated tunable single and double hole quantum dots based on core-shell Ge/Si NWs. Our measurements indicate that the hole g-factor in these systems is strongly suppressed and axially anisotropic, in agreement with theoretical expectations[18].

The authors wish to acknowledge helpful discussions with U. Zülicke (Massey University, New Zealand).

This work was supported by the Swedish SSF and VR, the Office of Naval Research (ONR), the Swiss SNF and Italian MUR (projects II04CBCF18 and RBIN045MNB).

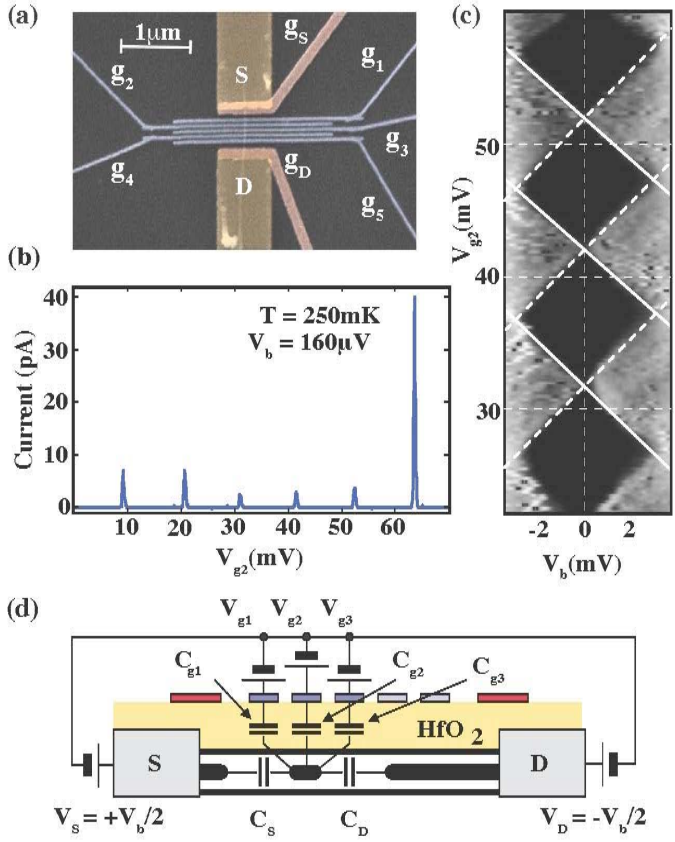

**Figure 1** (a) Scanning electron microscope image of one of the devices with the source and drain contacts colored yellow and the gates in blue and red. The thin Ge/Si NW can be seen as a white line extending vertically between the source and drain contacts. (b) Coulomb blockade oscillations in the current through the NW for a quantum dot confined between barrier electrodes $g_1$ ($V_{g1}$ = 0.63 V) and $g_3$ ($V_{g3}$ = 0.90 V) as a function of the plunger gate voltage $V_{g2}$. The average peak spacing is $\Delta V_{g2} \approx$ 10 mV. (c) Coulomb blockade diamonds in the differential conductance through the wire for the same single-dot configuration plotted using a logarithmic color scale and maximum conductance values of $0.1 e^2/h$. (d) Cross-sectional schematic of the device indicating the measurement set-up and capacitive coupling of the gates to the dot.

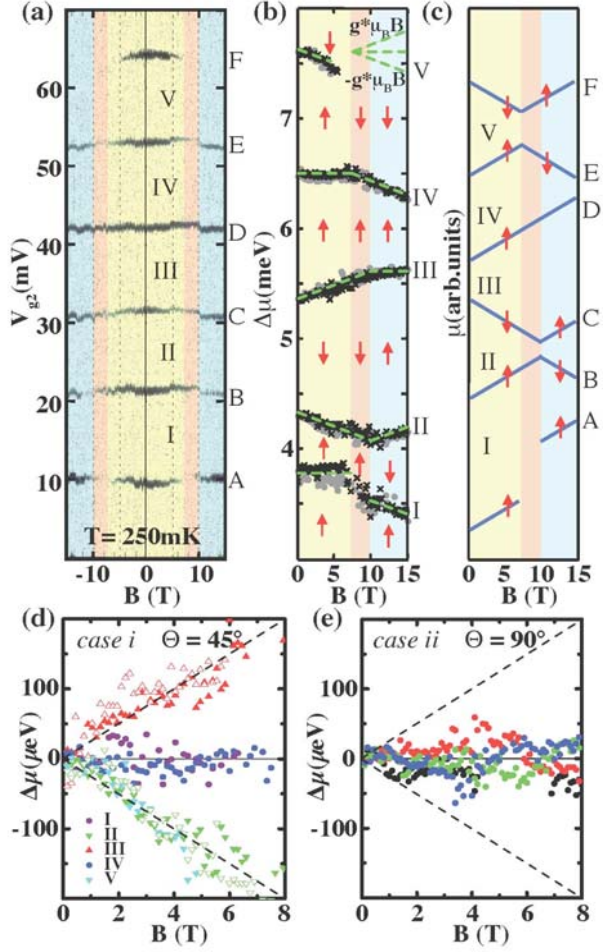

**Figure 2** (a) Parametric evolution of the Coulomb blockade peaks as a function of $B$. (b) Level separation $\Delta\mu(B)$ for positive (**x**) and negative (●) $B$. The curves were offset for better visibility and arrows indicate the spin filling of the dot assuming only three recurrent slopes occur (green dashed lines). (c) Schematic reconstruction of the spin filling sequence from (b). Two spin induced crossing between peaks B/C and E/F can be identified in the sequence of six peaks. (d+e) Evolution of the level separations for applied fields up to $B = 8$T, where (d) was taken with $B$ at a $45°$ angle to the wire axis and in (e) $B$ was applied perpendicular to the NW allowing us to estimate $|g_\parallel| = 0.60 \pm 0.03$ and $|g_\perp| < 0.12$.

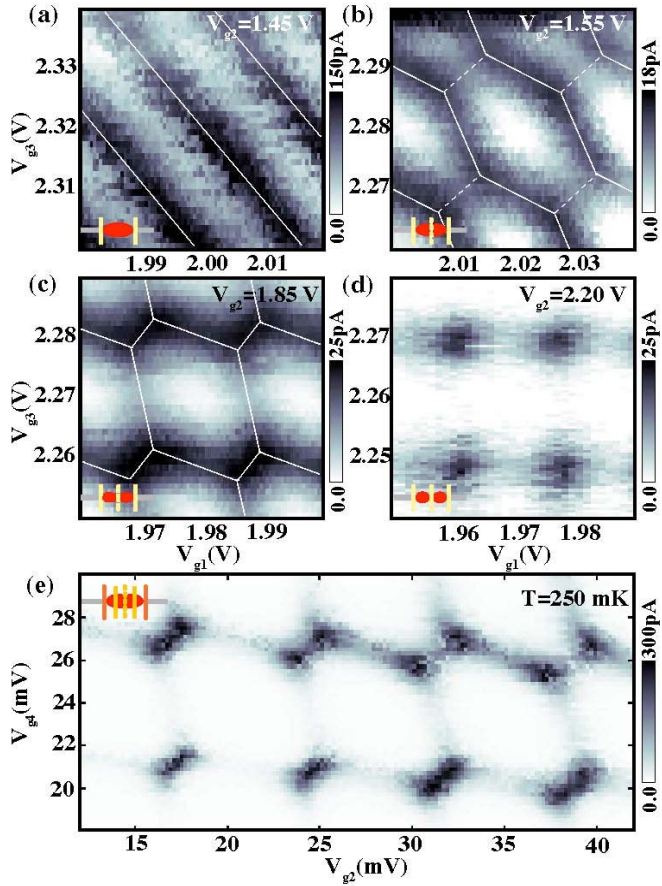

**Figure 3** (a)-(d) Double dot stability diagrams measured at T = 4.2 K as a function of the two barrier gate voltages $V_{g1}$ and $V_{g3}$ for increasing values of the centre gate voltage $V_{g2}$ (decreasing the coupling between the two dots). A linear current background was subtracted in the plots in order to maximize the contrast. (e) Alternate double-dot configuration using five gates to achieve independent control of the dot filling and the barrier transparencies at low temperature (T = 250 mK).